\documentclass[useAMS,usenatbib]{mn2e}
\usepackage{amssymb,amsmath,graphicx}
\usepackage{longtable}[1]
\usepackage{graphics}
\usepackage{threeparttable}
\usepackage{lscape}
\usepackage{color}
\title[Star Clusters in BCGs]{Probing cluster formation under extreme conditions: massive star clusters in blue compact galaxies}

\author[Adamo, Ostlin, \& Zackrisson]{A. Adamo$^{1}$\thanks{E-mail:
adamo@astro.su.se}, G. \"Ostlin$^{1}$, \& E. Zackrisson$^{1}$\\
$^{1}$Department of Astronomy, Stockholm University, Oscar Klein Center, AlbaNova, Stockholm SE-106 91, Sweden\\}
\newcommand{\araa}{ARA\&A}
\newcommand{\apj}{ApJ}
\newcommand{\aj}{AJ}
\newcommand{\mnras}{MNRAS}
\newcommand{\aap}{A\&A}

\newcommand{\nat}{Nature}

\newcommand{\apjl}{ApJ}
\newcommand{\apjs}{ApJS}
\newcommand{\msun}{M_{\odot}}
\newcommand{\ha}{H${\alpha}$}

\begin{document}

\date{Accepted 2011 July 4. Received 2011 July 4; in original form 2011 April 19}

\pagerange{\pageref{firstpage}--\pageref{lastpage}} \pubyear{2011}

\maketitle

\label{firstpage}

\begin{abstract}

The numerous and massive young star clusters in blue compact galaxies (BCGs) are used to investigate the properties of their hosts. We test whether BCGs follow claimed relations between cluster populations and their hosts, such as the the fraction of the total luminosity contributed by the clusters as function of the mean star formation rate density; the $V$ band luminosity of the brightest youngest cluster as related to the mean host star formation rate; and the cluster formation efficiency (i.e., the fraction of star formation happening in star clusters) versus the density of the SFR. We find that BCGs follow the trends, supporting a scenario where cluster formation and environmental properties of the host are correlated. They occupy, in all the diagrams, the regions of higher SFRs, as expected by the extreme nature of the starbursts operating in these systems. We find that the star clusters  contribute almost to the 20 \% of the UV luminosity of the hosts. We suggest that the BCG starburst environment has most likely  favoured the compression and collapse of the giant molecular clouds, enhancing the local star formation efficiency, so that massive clusters have been formed. The estimated cluster formation efficiency supports this scenario. BCGs have a cluster formation efficiency comparable to luminous IR galaxies and spiral starburst nuclei (the averaged value is $\sim 35$ \%) which is much higher than the  8-10 \% reported for quiescent spirals and dwarf star-forming galaxies. 

\end{abstract}


\section{Introduction}
 Luminous blue compact galaxies (BCGs) are star-forming systems with high specific star formation rate \citep{2001A&A...374..800O, 2005NewA...11..103S}. It has been suggested that BCGs may have accounted for $\sim$ 40\% of the total star formation rate (SFR) density at redshift $0.4<$z$<1.0$ playing an important role in the star formation history (SFH) of the Universe \citep{1997ApJ...489..559G}.  However, in the local Universe, their contribution has dropped drastically and  luminous BCGs with high SFR have become rare objects \citep{1997ApJ...489..559G, 2004ApJ...617.1004W}. At a distance of 100 Mpc (z$\sim$0.03), we count only a handful number of BCGs with a SFR higher or close to 5 $\msun$yr$^{-1}$.  Among the systems included in this work, Haro 11 \citep{A2010}, ESO 185-IG13 \citep{A2010c}, and Mrk\,930 \citep{A2011a} are starburst BCGs with SFRs exceptionally high for local irregular galaxies. Local luminous BCGs display physical conditions (low metallicity and dust content), morphologies (compactness of the starburst regions), and feedback mechanisms \citep[e.g.][]{2009AJ....138..923O} similar to their high redshift counterparts. 

Numerical simulations and theoretical arguments based on the Lambda Cold Dark Matter model of the Universe predict that smaller galaxies formed first and were then accreted into more massive systems  \citep[hierarchical growth,][]{2000MNRAS.319..168C, 2005Natur.435..629S, 2005ApJ...631..101P}. Therefore, it is expected that the primordial galaxies were chemically unevolved dwarf galaxies, the so-called "building-blocks" systems, which accrete and merge into more massive units. Recent observations of galaxies at very high redshifts  (z$\sim7$, i.e. at the reionization epoch) have revealed  compact single or double nuclei systems with extended nebular features \citep{2010ApJ...709L..21O}. Some of these z$\sim7$ objects have also  been detected at longer IR wavelengths. These data have been used to constrain and study the spectral energy distributions (SEDs) of these high-z systems. The inferred SFRs are between 5 to 20 $\msun$yr$^{-1}$ \citep{2010ApJ...713..115G}. In general, they have estimated stellar masses of $10^{8-9}$ $\msun$ and fainter UV luminosities than lower redshift Lyman break galaxies \citep{2010ApJ...719.1250F}, i.e. their properties resemble those of the luminous BCGs.

The studies of BCGs can, therefore, give important insights in understanding how star formation proceeds in dwarf starburst galaxies and possibly in high-redshift systems, with spatial and spectral resolution impossible to achieve for the latter with the current facilities. High-resolution imaging data of BCGs revealed that the starburst regions in these galaxies are formed by massive and young star cluster complexes \citep[e.g.][]{2003A&A...408..887O, A2010, A2010c, A2011a}. The peak age of the star cluster distributions and the estimated ages of the starbursts in these systems are in good agreement. The host morphologies suggest that the galaxies have recently undergone a merger or interaction event, which has likely refurbished the galaxy with  metal-poor gas, and triggered a vigorous starburst episode. 

Star clusters are a natural outcome of the star formation process (see \citealp{2003ARA&A..41...57L} for a review). However, for very young stellar systems (e.g. a few Myr or less), it is not trivial to make a clear definition of a cluster \citep{2010MNRAS.409L..54B}. Usually, it is assumed that stars form in a clustered fashion and that, after roughly 10 Myr, 90 \% of clusters are destroyed due to  gas expulsion.  \citet{2010MNRAS.409L..54B}  suggest that all the stars form in a continuos hierarchy, even the dense regions and estimate that, in the solar neighbourhood,  only $\sim 26 \%$ of young stellar objects (new born stars) are located in denser regions, i.e. embedded star clusters. The remaining stars form in associations and agglomerates following a hierarchical continuum distribution with clusters at the bottom of the process. \citet{2010ARA&A..48..431P} and \citet{2010MNRAS.tmpL.168G} defined an empirical relation to separate clusters from loose associations. For the latter the crossing time is much larger than the age of the stars, i.e. they are dynamically unbound systems. They observed that the division between bound and unbound systems becomes more clear once the gas-expulsion phase is over ($\sim$ 10-20 Myr). 

Cluster formation at high-redshift is almost unknown. In the local Universe, we observe the evolved counterparts, i.e., the globular clusters (GCs). However, there is no direct evidence that the young star clusters we observe locally will eventually evolve into GCs \citep{2007ChJAA...7..155D}.  If one looks at the number of GCs per mass (luminosity) bins (i.e., $dN(M)/dM=CM^{\eta}$) the distribution can be fitted by two power-laws with a slope of $\sim-2.0$ at the massive (luminous) ends, a break corresponding to the $M_c\sim2\times10^5 \msun$, and a flattening ($\eta\sim-0.2$ which variates from system to system) at lower masses. Several possible scenarios are addressed by theoretical studies to explain the origin of the GC mass (luminosity) function: dynamical evolution, i.e. due to preferential disruption of the low mass systems; or  a primordial origin, i.e. the GC mass function has been established at the time when the GCs formed \citep{2007ChJAA...7..155D}. Cosmological simulations seem to suggest that GCs may have formed in dark matter halos \citep[e.g.][]{2005ApJ...623..650K, 2005ApJ...619..258M}. If this is the case, there is no connection between the YSCs forming at the present time and the ancient GCs. However, some recent works \citep[see][for a review on GCs]{2006ARA&A..44..193B} suggest that cluster formation in dwarf galaxies may be the key to explain observed properties of the ancient GCs (blue versus red, i.e. metal poor versus metal rich). \citet{2010ApJ...718.1266M} recovered a bimodal metallicity distribution as a product of cluster formation in different phases of galaxy evolution. Cluster formation in dwarf galaxies produced the blue, metal-poor GCs.  Dwarf systems were successively accreted to form more massive systems. During the merger and formation of these massive galaxies the more metal-rich clusters were formed. Accretion of dwarf galaxies together with their GC systems is also one of the proposed scenarios by \citet{2011A&A...525A..20C} to explain why the younger GCs in S0 type of galaxies appear blue instead of the expected metal-rich populations \citep{2006ARA&A..44..193B}.

In the present work, we will investigate how cluster formation has proceeded in BCGs. We will test whether BCGs follow the cluster-host relations available in the literature and constrained using local star-forming galaxies, like spirals and dwarfs. These results will be used to constrain the environmental properties of BCGs and whether star formation operates on similar modes even under extreme conditions. 

The paper is organized as follow: In Section 2, we present the 5 BCG targets used in this work. In Section 3, we first discuss the uncertainties which affect the analysis. In the second part of this section, we show the three cluster-host relations including the BCG sample. A discussion of the results is presented in Section 4. Here, we also discuss possible similarities between BCGs and high redshift galaxies. Conclusions are summarized in the last section.

\section[]{The data}
\label{data-sample}

In this section we shortly introduce the BCGs included in the analysis. Some of these BCGs have been studied in a series of 3 papers: Haro 11 analysis is presented in \citet[][]{A2010}; ESO 185-IG13 (ESO 185) in \citet[][]{A2010c}; and Mrk\,930 in \citet[][]{A2011a}. We refer to those papers for details on the analysis of the data used in this work.

\subsection{ESO 338-IG04}

\begin{figure}
\resizebox{\hsize}{!}{\rotatebox{0}{\includegraphics{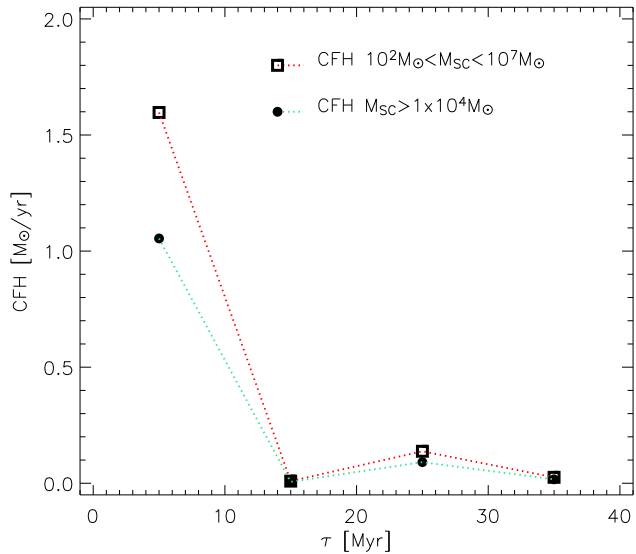}}}
\caption{Cluster formation rates during the last 40 Myr of starburst activity in ESO 338. The filled dots connected by the green dotted line show the observed CFR derived from clusters more massive than $10^4 \msun$. The squares show the derived CFR, if the total mass in clusters less massive than $10^4 \msun$ is extrapolated using a CMF with index $-2.0$ down to $10^2 \msun$. The data used are from \citet{2003A&A...408..887O}.}
\label{CMF_eso338}
\end{figure}

The analysis of the star cluster population in ESO 338-IG04 (ESO 338) has been presented in \citet{2003A&A...408..887O}. The masses have been obtained from models that assume a Salpeter initial mass function \citep[IMF,][]{1955ApJ...121..161S}. We show, in Figure~\ref{CMF_eso338}, the cluster formation history during the last 40 Myr of galaxy evolution. Using the age and mass estimates from \citet{2003A&A...408..887O}, we assumed that the analysis is complete in detecting clusters more massive than $5\times10^3$ $\msun$ formed during the last 40 Myr. A power law cluster mass function with index $-2.0$ has then been used to extrapolate the total fraction of mass in clusters including objects with masses between $10^2\leq$M$\leq5\times10^3 \msun$. Following \citet[][]{2004MNRAS.350.1503W}, we assume for simplicity that a cluster population forms every 10 Myr and estimate the cluster formation rate (CFR) in the galaxy. In agreement with \citet{2003A&A...408..887O}, we observe a cluster formation enhancement between 20 and 30 Myr ago. However, the peak of cluster formation is younger than 10 Myr, similarly to what is found in Haro 11, ESO 185, and Mrk\,930. ESO 338 is forming roughly 1.6 $\msun$yr$^{-1}$ of stars in clusters. 

We defined the cluster formation efficiency (CFE, indicated hereafter also as $\Gamma$) as that the fraction of stars formed in star clusters. If we compare the CFR to the mean SFR happening in the galaxy, we obtain the CFE$=$CFR/SFR \citep[see G10 and][hereafter B08]{2008MNRAS.390..759B}. Using a SFR of 3.2 $\msun$yr$^{-1}$ \citep{2001A&A...374..800O}, we find that $\Gamma \sim$50 $\pm$ 10 \%, i.e. 50 \% of  the ESO 338 stars are formed in clusters. 

\subsection{SBS 0335-052E}

The nuclear region of this extreme metal-poor galaxy is dominated by 6 massive young star clusters which have been referred to in the literature as super star clusters (SSCs). These SSCs have been extensively studied before (e.g., \citealp{2008AJ....136.1415R}; \citealp{2009ApJ...691.1068T}; \citealp{A2010b} among the most recent published works on this subject).

\begin{figure}
\resizebox{\hsize}{!}{\rotatebox{0}{\includegraphics{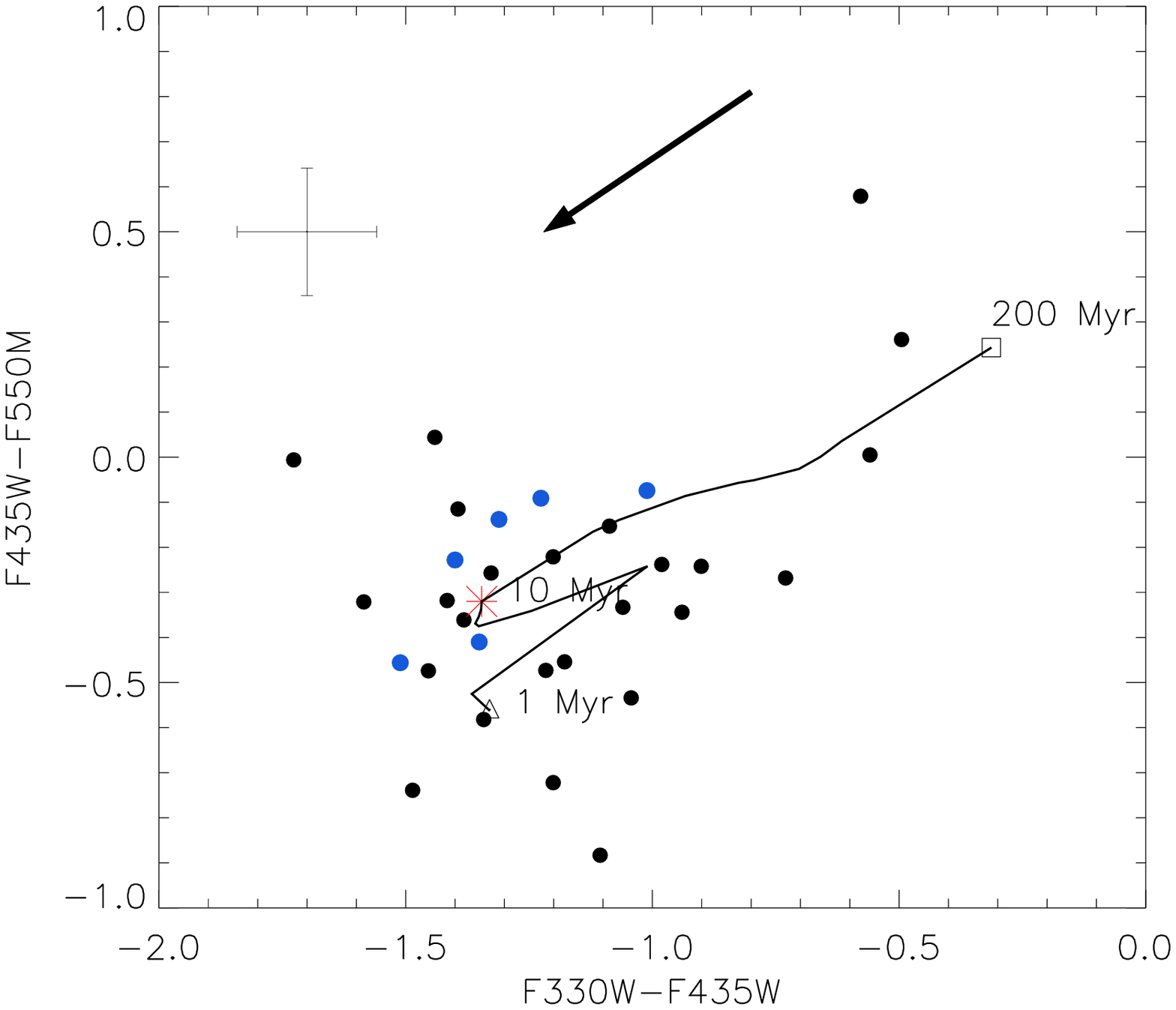}}}
\resizebox{\hsize}{!}{\rotatebox{0}{\includegraphics{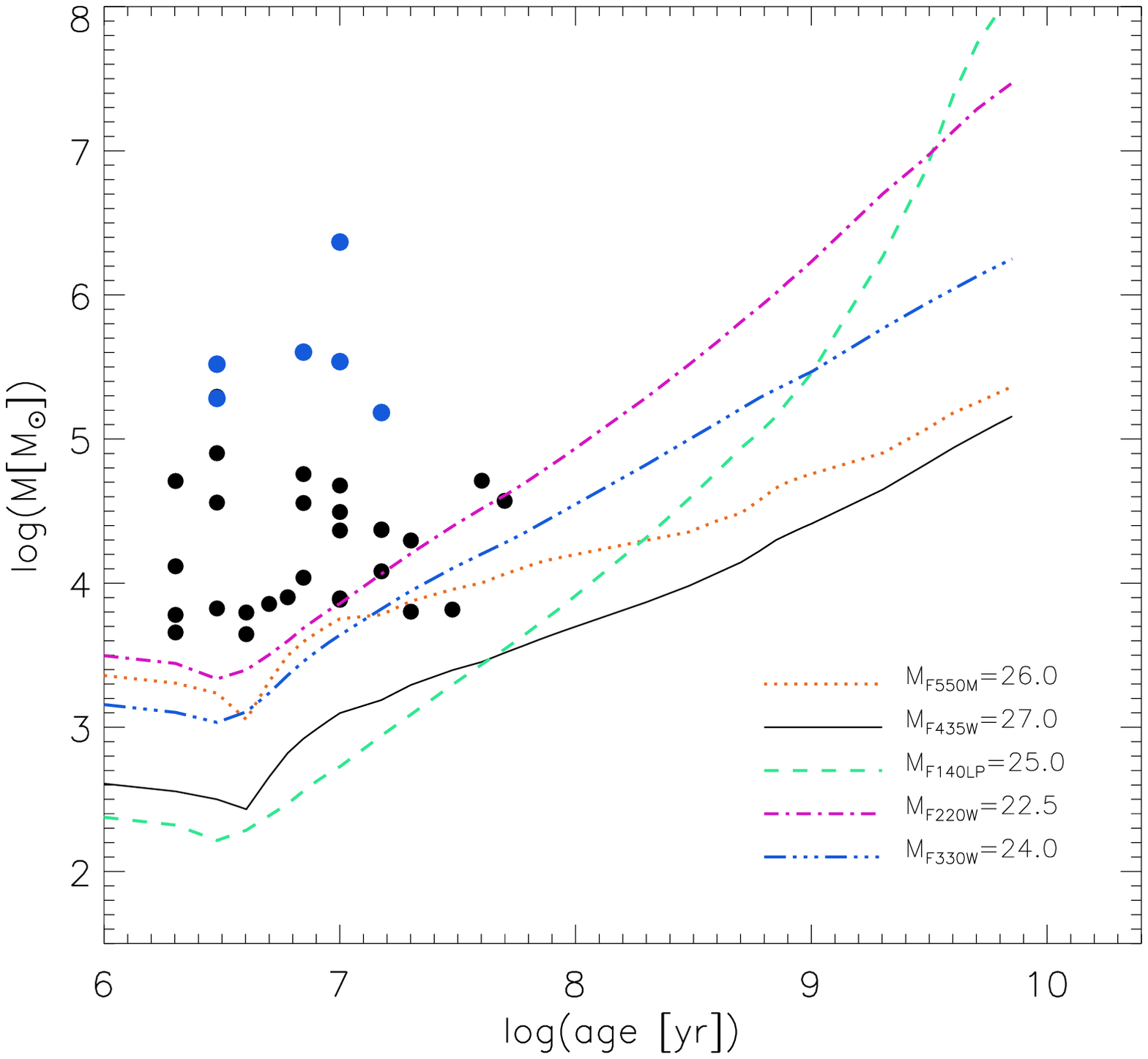}}}
\caption{In the top panel the color-color diagram of the whole cluster population detected in SBS 0335-052E is shown. Blue dots  indicate the six super massive star clusters, which dominate the starburst in the galaxy and have already been studied in several works (e.g. \citealp{A2010b} for references). The black dots represent the newly studied lower mass cluster population. The evolutionary track has a metallicity of $Z = 0.0004$ \citet[see][for details]{A2010b}. Some important ages are labelled. The extinction vector indicates in which direction the clusters move if a de-reddening of  A$_V=1$ mag is applied to the observed colors. A mean photometric error-bar is shown. In the bottom panel, we show the mass-age diagram of the star cluster population. The several lines indicate the detection limits reached in our observed data (shown in the inset).  See main text for details.}
\label{mas-age}
\end{figure}

In the present work we explore the lower mass cluster population of the galaxy. Part of this underlying cluster population has already been revealed by \citet{1998A&A...338...43P}.  We have access to $FUV$ ({\it SBC}/F140LP from GO 9470, PI: Kunth) and optical images ({\it ACS} F220W, F330W, F435W, F550M from GO 10575, PI: Ostlin) for the galaxy. The reduction of the science frames is described in \citet{2009AJ....138..923O}. 
The extraction of the cluster candidates has been done using the {\tt PyRAF} package {\tt DAOFIND} on the $B$ band frame (F435W). The catalogue has been cleaned by eye of all the detections which didn't show a clear visual counterpart.  With this first catalogue, we have done photometry on all the frames from the $FUV$ to the $V$ (F550M) bands. Applying the same method \citep[see][]{A2010, A2010c, A2011a} we have included in the final catalogue only cluster candidates with detection in at least 3 filters and a photometric error $\sigma_m < 0.2$ mag. The photometric properties of the final cluster population are shown in the color-color diagram in Figure~\ref{mas-age}. The filter combination corresponds to a $U-B$ (F330W$-$F435W) versus a $B-V$ (F435W$-$F550M) color. The colors are not dereddened  but clearly suggest that the cluster candidates are not older than $\sim 100$ Myr.

The spectral energy distribution (SED) fitting procedure is  described in \cite{A2010}. The used models are presented in \citet{A2010b}. The output age and mass of the clusters are shown in the mass-age diagram in Figure~\ref{mas-age}. In blue dots, we show the 6 SSCs previously analysed. The underlying black dots represent the low mass cluster population that is present in the galaxy, detectable at the detection limits imposed by the data. The masses are smaller than $\sim 5\times 10^4$  $\msun$ and the ages are younger than 50 Myr. It is still under debate whether an old stellar population (older than 100 Myr) exists in SBS 0335-052E \citep{2001A&A...371..429O} or if this galaxy has recently formed \citep{1998A&A...338...43P}. In previous star cluster analyses of other BCGs (ESO 338, Haro 11, ESO 185, Mrk\,930), we have found a trace of some old GCs, supporting the evidence of an old underlying stellar population in those galaxies.  In the case of  SBS 0335-052E, however, the non detection of massive GCs cannot prove/disprove either of the two proposed scenarios. We observe that, because of the detection limits (see inset in Figure~\ref{mas-age}), our analysis is limited to GCs with masses higher than $5\times 10^4$ $\msun$ between 100 Myr and 1 Gyr and even more massive at older ages. Therefore, we cannot exclude that this galaxy has low-mass GCs.

Using the same method as in the case of ESO 338, we infer  the mass in clusters formed during the last 10 Myr in SBS 0335. We observe that the total mass contained in detected clusters younger than 10 Myr is  $1.95\times 10^6$ $\msun$. The mass contained in the 5 SSCs (one of them is much older, e.g. from H$\alpha$ equivalent width, the age is constrained  to $\sim13$ Myr is constrained, see \citealp{A2010b}) is roughly 73 \% of this total mass ($1.42\times 10^6$ $\msun$). Assuming a power law cluster mass function and that we are complete in detecting clusters more massive than $5\times10^3 \msun$, we estimate a total mass in clusters younger  than 10 Myr of $6.4\times10^6 \msun$. The observed CFR in systems more massive than M$>5\times10^3$ $\msun$ is 0.2 $\msun$yr$^{-1}$, while the extrapolated CFR (M$>10^2$ $\msun$)$=0.64$ $\msun$yr$^{-1}$. The $\Gamma$ value of SBS 0335 is  49 $\pm 12 \%$, using a SFR of 1.3 $\msun$yr$^{-1}$. These values have been estimated after that a correction for Salpeter IMF has been applied (see Section~\ref{sfr_unc}).

\section{Relations between cluster and star formation rate}

The formation of a cluster appears to be correlated with the properties of the host galaxy. A common observation is that galaxy mergers produce more numerous and more massive clusters than quiescent spirals \citep[][and references therein]{2009A&A...494..539L}. {\it Sampling statistics}, known also as  size-of-sample effect, (i.e., galaxies with a more numerous cluster population have higher chances to sample the cluster mass function, CMF, at higher masses) is a possible explanation for this trend \citep[][hereafter L02]{2000astro.ph.12546W, 2002AJ....124.1393L}. On the other hand, the host environment can play its role in determining the mass of the forming clusters \citep{2006A&A...450..129G}. Numerical simulations of different host environments suggest that the shear in rotationally supported galaxies (i.e., spirals) acts on the collapse of the giant molecular clouds (GMCs), causing fragmentation and favouring the formation of the less clustered OB associations and low mass clusters  \citep{2010ApJ...724.1503W}. The lack of rotation in dwarf galaxies and high external pressures in merging systems favour the collapse of  massive and gravitationally bound cluster. These two scenarios were also addressed by \citet{2002AJ....123.1454B} to understand cluster formation in dwarf galaxies. They studied the star cluster populations of nearby dwarf galaxies and observed that not all the systems had bound and luminous clusters. However, some of them hosted one or a few very massive ones. They suggested that, with respect to spiral galaxies which form more clusters and can, therefore, sample the cluster mass function homogeneously up to high mass bins, the cluster formation in dwarf galaxies is possibly dominated by stochasticity together with favourable physical conditions to form single massive clusters. 

Observed empirical relations between the properties of  the young star clusters and the SFR in the host support the size-of-sample effect scenario.  \citet[][hereafter LR00]{2000A&A...354..836L} first noticed that the fraction of luminosity contained in the young star clusters and the SFR of the host galaxy are correlated (T$_L$(U)-$\Sigma_\mathrm{SFR}$ realtion), i.e., higher SFRs correspond to a more numerous cluster population (higher cluster formation efficiency).  In a follow-up work, L02 found evidence of a positive correlation between the visual  luminosity of the brightest star cluster and the SFR in the host ($M_V^{\textnormal{brightest}}$-SFR relation). The relation between the two quantities can be understood if higher SFRs enabled the formation of more massive clusters. B08 enlarged the sample of L02, including resolved close-by star-forming regions and luminous and ultra-luminous IR galaxies (LIRGs and ULIRGs). He noticed that the $M_V^{\textnormal{brightest}}$-SFR relation holds over several orders of magnitude in SFR values, suggesting that the youngest brightest cluster is a fairly good indicator of the present SFR in the galaxy.  These observed relations clearly point towards a scenario where the cluster formation is intimately  correlated with the star formation process, or in other words, the birth of a cluster is a product of an universal star formation process which operates on many scales of efficiency. 

 In a recent work by G10, it has been inferred that the present CFE (clusters formed in the last 10 Myr) is higher for a higher current SFR in the host, the so-called $\Gamma$-$\log(\Sigma_\mathrm{SFR})$ relation. \citet{2011arXiv1101.4021S} used a sample of 5 nearby spiral galaxies  to test the G10 relation using two different methods to estimate the CFRs. They observed that the recovered data points, in spite of the method used, scattered around  the expected values and were impossible to reconcile with the Goddard et al. relation (see Figure~\ref{sfr-cfr}). Despite the discrepancy between the two results, we will include the $\Gamma$-$\log(\Sigma_\mathrm{SFR})$ relation in the tests we will perform for the BCGs. A discussion of the uncertainties affecting this relation will be presented in the next section.  They may  explain the disagreement between Goddard et al. and Silva-Villa \& Larsen results.
 
To investigate whether BCGs follow these cluster-host relations we use the quantities listed in Table~\ref{table-obs} and \ref{table-obs2}. Before we test the relation and compare our data to the ones published in the literature we discuss, briefly, the main source of uncertainties associated with the derived parameters and the used methods.

\begin{table*}
  \caption{For each BCG, several quantities are listed in the table. From the left to the right column: name of the target; distance, absolute magnitude M$_V^{\textnormal{brightest}}$ of the brightest young cluster in the galaxy; radius containing 80 \% of the total $B$ luminosity of the galaxy, R$_{80\%}$; mean SFR; surface density of SFR; CFR; and $\Gamma$. The SFR and CFR are estimated assuming a Salpeter IMF (see Section~\ref{sfr_unc}).}
\centering
\begin{threeparttable}
  \begin{tabular}{|c|c|c|c|c|c|c|c|}%{@{}llll|@{}}
  \hline
 target & distance\tnote{h}& $M_V^{\textnormal{brightest}}$ & R$_{80\%}$ & SFR & $\Sigma_\mathrm{SFR}$ & CFR & $\Gamma$ \\
   \hline
&Mpc & mag & (Kpc)& ($\msun$yr$^{-1}$)&($\msun$yr$^{-1}$Kpc$^{-2}$) &($\msun$yr$^{-1}$) & (\%)\\
   \hline
 ESO 338-IG04&37.5&-15.5\tnote{a}& 0.8 &  3.2\tnote{b}	&1.55 &     1.6  & 50$\pm 10$\\

 Haro 11 &82.3 & -16.16\tnote{c} & 1.8 &	22.0\tnote{c}	&2.16& 11.2& 50 $^{+13}_{-15}$\tnote{c}\\

 ESO 185-IG13&76.3 &-14.55\tnote{d}&  1.98&6.4\tnote{d}&0.52& 1.7\tnote{d} & 26 $\pm 5$\tnote{d}\\

 Mrk\,930&71.4 &-15.17\tnote{e}&1.7&	5.34\tnote{e}	&0.59 & 1.33\tnote{e} & 25 $\pm 10$\tnote{e}\\

 SBS 0335-052E\tnote{f}&54&-14.28&0.66&1.3&0.95& 0.64 (0.14)\tnote{g} & 49 (10)\tnote{g} $\pm 15$ \\
 \hline
\end{tabular}
 \begin{tablenotes}
 \item[a] \citet{2003A&A...408..887O}; \item[b]  \citet{2001A&A...374..800O};
  \item[c] \citet{A2010};
 \item[d] \citet{A2010c};
 \item[e] \citet{A2011a};
 \item[f] \citet{2008AJ....136.1415R};
 \item[g] The values indicated between brackets are estimated for the 5 SSCs with ages $\leq 10$ Myr (see Figure~\ref{mas-age} and \citealp{A2010b}). However, in the analysis we use the values obtained including the whole cluster population.
 \item[h] data from NED, http://ned.ipac.caltech.edu/
 \end{tablenotes}
 \end{threeparttable}
\label{table-obs}
\end{table*}

\begin{table}
  \caption{From the left to the right column: name of the target; extended radius, R$_G$, values in arcsec are in brackets; the specific luminosity, T$_{\textnormal{L}}$, of the cluster population with respect the host in $FUV$ (central wavelengths of the filter $\sim$ 0.14 $\mu$m), $U$, and $B$. In the case of SBS 0335-052E, we include between brackets the recovered T$_{\textnormal{L}}$ if only the 6 SSCs are considered.}
\centering
  \begin{tabular}{|c|c|c|c|c|}%{@{}llll|@{}}
  \hline
 target & R$_G$(Kpc) & T$_{\textnormal{L}}$(UV) &  T$_{\textnormal{L}}$(U) &T$_{\textnormal{L}}$(B) \\

   \hline
 ESO 338&1.4 (7.9")& -	& 34.6&20.9\\

 Haro 11 & 4.9(12.5") & 20.3& 11.8& 16.4 \\

 ESO 185& 4.9 (13.5")& 14.3&9.8&7.1\\

 Mrk\,930& 4.5 (12.5") &19.2&22.3&12.1\\

 SBS 0335& 2.6 (10.0") &29.4(21.3)&32.8(24.0)&38.4(26.3) \\
 \hline
\end{tabular}
\label{table-obs2}
\end{table}

\subsection{Uncertainties}
\label{area}

\subsubsection{The estimate of $\Sigma_\mathrm{SFR}$} 
\label{sfr_unc}
The estimate of $\Sigma_\mathrm{SFR}$ depends on the SFR in the galaxy and the size of the host. The SFR is an averaged value of the star formation happening in the system. Many tracers are used in the literature to estimate the SFR and they do not always produce the same results. Moreover, in a galaxy there are regions which are quiescent and others starbursting. To estimate a meaningful $\Sigma_\mathrm{SFR}$ value, it is hence necessary to determine the size of the region which is currently producing stars. Since there is not a standard way to estimate the areas, the $\Sigma_\mathrm{SFR}$ can be diluted or overestimated. When SFR or $\Sigma_\mathrm{SFR}$ are used to compare properties of sample of galaxies it is necessary to keep in mind that some of the scatter is due to the different methods applied to estimate these quantities. In the present analysis, we will compare our sample of BCGs with other literature data. For the BCGs, the SFRs are obtained from measurements of the \ha \, fluxes \citep[applying the][law]{1998ARA&A..36..189K}, and the sizes are estimated in a homogeneous way. 

Finally, as already pointed out by G10, the Kennicutt law for SFR has been calibrated assuming a Salpeter IMF. However, when $\Gamma$ is estimated, we compare CFR and SFR in the same host. Since in our star cluster analysis we have assumed a Kroupa IMF (2001), we have, for sake of consistency,  applied a conversion factor in order to obtain cluster masses for a Salpeter IMF. The CFR and $\Gamma$ values listed in Table~\ref{table-obs} are for cluster masses derived assuming a Salpeter IMF. 
 
\subsubsection{The area of the starburst}
The outskirts of BCGs are quite extended, while the starburst regions are confined to the central areas of the galaxies.    In the current analysis we use two different sizes (radii): one which includes the starburst and outskirts (R$_G$, radius of the galaxy); the other to delimit the area where the star-forming regions are contained (R$_{80\%}$, radius of the star-forming region). The R$_G$ is the most extended radius we can infer from our data and  is used to estimate the luminosity of the targets, L$_{\textnormal{host}}$, in a few bands (see below). The R$_{80\%}$ is used for constraining the density of SFR in the galaxy, $\Sigma_\mathrm{SFR}$. The latter is a measure of the rate at which the star-forming regions are producing stars. Therefore, it is important to use a homogeneous method to estimate the area of these active regions when different galaxies are compared.
\begin{figure}
\resizebox{\hsize}{!}{\rotatebox{0}{\includegraphics{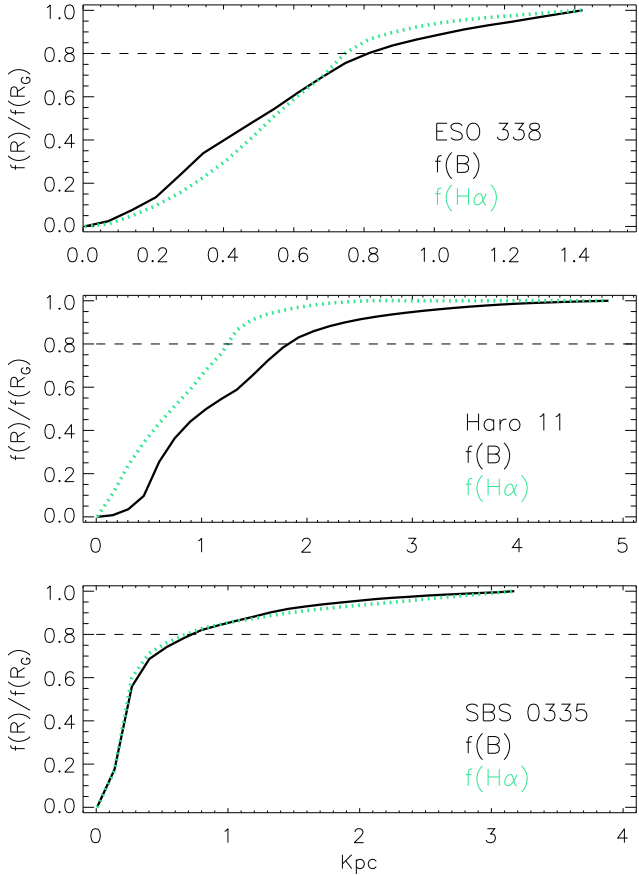}}}
\caption{The fraction of galactic luminosity contained inside increasing radii as function of the total luminosity at the most extended galactic radius, R$_G$. In the inset we show which target is analysed. The fraction of flux in $B$ band, f($B$) is showed as a black solid line, whereas the F(\ha) is the green (grey) dotted line.}
\label{grow}
\end{figure}

 To obtain an estimate of  R$_{80\%}$ radius, we have looked at the fraction of galactic flux contained in growing aperture radii as function of the total flux contained inside  R$_G$. A direct check on the frames shows that the R$_{80\%}$ encloses generously the starburst regions in all the BCGs here studied. As a further check we have compared (Figure~\ref{grow}) the fraction of flux in the $B$ band and  in \ha \ for 3 of the 5 BCGs (we refer to \citealp{2009AJ....138..923O} for the reduction and analysis). \ha \ emission is usually considered a standard tracer of the SFR \citep{1998ARA&A..36..189K}. On the other hand, the $B$ band flux is dominated by the light produced by young stars and can also be considered a reliable indicator of the location of the starburst regions.  We compare the fraction of growing flux in the two tracers to check whether R$_{80\%}$ can be considered a good estimate of the size of the starburst regions in the galaxy (see Figure~\ref{grow}).

In the central regions, where the starburst dominate, the \ha \ luminosity distribution varies in the 3 galaxies. In ESO 338 (top panel, Figure~\ref{grow}), we observe that the \ha \  luminosity is less centrally concentrated than the $B$ band luminosity. However, at R$_{80\%}$, we observe an opposite behaviour and roughly 90 \% of the \ha \ flux is enclosed. This effect is caused by the presence of a very massive and young cluster in the centre of ESO 338 \citep{2007A&A...461..471O}, which has cleaned the surrounding regions of the nebular gas (see image of the galaxy in \citealp{2009AJ....138..923O}), causing a dearth of \ha \, emission. In Haro 11 (center panel, Figure~\ref{grow}), the situation is inverted. The \ha \ luminosity is contained in a much smaller region than the $B$ band one and, at R$_{80\%}$, we detected almost the 100 \% of \ha \ flux produce by the galaxy. Finally, in the case of SBS 0335-052E (bottom panel), we see that \ha \ and $B$ band luminosities grow in a similar way. 

In all the cases within the uncertainties, R$_{80\%}$, estimated from the $B$ band, incorporates or is in a fairly good agreement with \ha \, luminosity distribution. Therefore we use R$_{80\%}$ to estimate the area of the star-forming regions in the galaxies and, thus, the $\Sigma_\mathrm{SFR}$.

\subsubsection{The distinction between bound clusters and associations} 

The distinction between clusters and associations has not been applied to our BCGs sample, neither to the data used to obtain the three relations discussed in this paper. Two of these relations, the T$_L$(U)-$\Sigma_\mathrm{SFR}$ and the $\Gamma$-$\log(\Sigma_\mathrm{SFR})$, can be drastically affected. If the real number of clusters is overestimated (e.g. stars in unbound systems are counted as clusters and not as field stars) it will alter both quantities, T$_L$(U) and $\Gamma$. B08 refers to the  M$_V^{\textnormal{brightest}}$-SFR relation as an evidence of an universal cluster formation efficiency, $\Gamma$ of 8 \% if clusters form with a Schechter CMF. This relation also implies  that galaxies with higher SFR form more massive clusters due to a statistical sampling effect. However, if the cluster formation efficiency is constant for increasing SFR,  the  $\Gamma$ and the T$_L$(U) should also be constant. Therefore, the positive relations of  T$_L$(U) and $\Gamma$ versus SFR could be caused by a contamination of unbound systems. There is no other evidence, which proves a universally constant cluster formation efficiency, nor is there an accessible way to estimate whether the T$_L$(U)-$\Sigma_\mathrm{SFR}$ and the $\Gamma$-$\log(\Sigma_\mathrm{SFR})$ relations still hold after a re-analysis including only bound objects. In this paper, we assume  that clusters and associations are a product of the same star formation process, happening under different physical conditions and at different scales. This assumption is also supported by the distribution presented in \citet{2010MNRAS.tmpL.168G}. They see no clear distinction between associations and clusters during the first 10 Myr of their formation but a continuous distribution. Only when these systems age is there a clear distinction with only clusters remaining tightly bound \citep[see Fig. 2 by][]{2010MNRAS.tmpL.168G}.  We refer to a cluster or association as a clustered structure, knowing that only bound clusters will survive longer \citep{2006MNRAS.369L...9B} and, eventually, become globular clusters. Since we are not able - with the current data - to disentangle these uncertainties, we will limit our analysis to compare the properties of the BCGs with these known relations and discuss possible implications, assuming that the estimated $\Gamma$ and T$_L$(U) are, indeed, upper limits to the real values.

\subsubsection{Distance of the targets} 

\begin{figure}
\resizebox{\hsize}{!}{\rotatebox{0}{\includegraphics{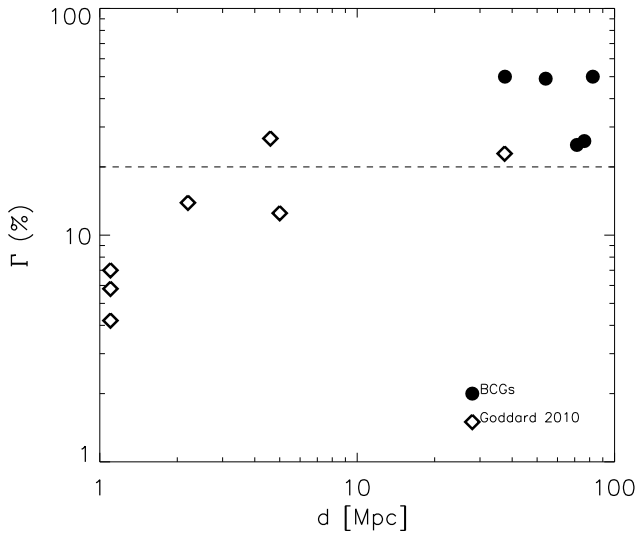}}}
\caption{The distance of the G10 sample (diamonds) and of the BCGs (dots) are plotted as function of $\Gamma$. The three diamonds at 1 Mpc are the two Magellanic Clouds and the Milky Way. Their distances have been plotted at 1 Mpc to make the plot more clear. A dashed line separates systems with CFE above 20 \%}
\label{dist}
\end{figure}

 In general, the distance of the host system can affect the quality of the cluster analysis. For increasing distance, our ability to resolve single clusters diminishes rapidly. Blending and crowding (clusters usually form in complexes) can significantly produce overestimates of the quantities we are interested in, T$_L$(U), M$_V^{\textnormal{brightest}}$, CFR and, thus, $\Gamma$. The cluster analysis of 3 of 5 BCGs are likely affected by blending, i.e., the most distant ones (distances are listed in Table~\ref{table-obs}): Haro 11, ESO 185, Mrk\,930. The other two BCGs are close enough to resolve most of the clusters. Since derived cluster properties can be overestimated because of blending, we tried to look for any positive correlation between distance and $\Gamma$. In Figure~\ref{dist}, the distance of the hosts of the G10 sample and of the BCGs are plotted as function of their $\Gamma$. There is a trend between a high value of $\Gamma$ and farther galaxies. However, number statistics are very poor and an observational bias affects this outcome. Low SFR systems are easier to observe locally, while their detection efficiency decreases rapidly at larger distances. On the other hand, high SFR hosts are easily observed at larger distances because of their high luminosities. Another element that needs to be considered is that environments with SFRs comparable to the BCG ones are rare in the local volume ($<$ 20 Mpc). In this case, the position of the M83 starburst nucleus at about 5 Mpc and $\Gamma > 20$ \% gives evidence that high star formation efficiency are also observed in systems where blending and crowding are not an issue. If we restrict the comparison to the BCG sample, we do not see any difference between the distant and nearby systems, supporting the idea that blending is not severe in either of the 3 most distant galaxies.

\subsection{The fraction of light in star clusters}

\begin{figure}
\resizebox{\hsize}{!}{\rotatebox{0}{\includegraphics{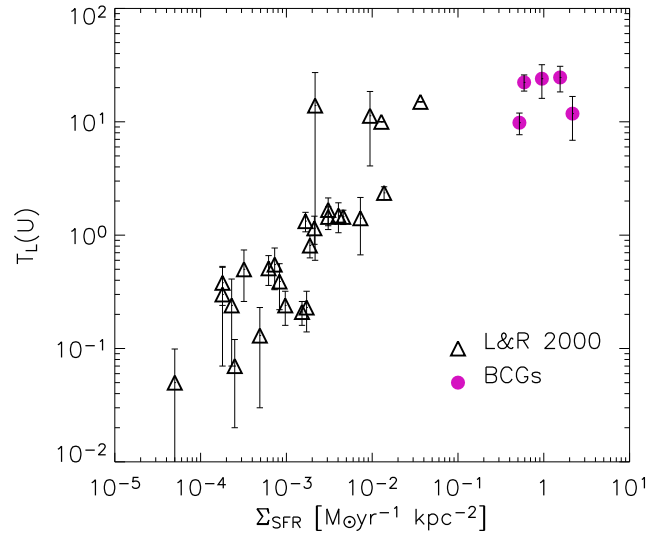}}}
\caption{T$_L$(U)-$\Sigma_\mathrm{SFR}$ relation by LR00. The sample used by LR00 is shown with black triangles. The BCGs are added as purple dots.}
\label{tlu-sfr}
\end{figure}

LR00 introduced the specific luminosity for young star clusters defined as T$_{\textnormal{L}}=100L_{\textnormal{clusters}}/L_{\textnormal{host}}$. T$_{\textnormal{L}}$ gives an estimate of the fraction of the total galaxy light that is produced by stars in clustered regions. Using a sample of galaxies which includes quiescent spirals and star-forming dwarf systems, LR00 found that the specific luminosity in the $U$ band, T$_{\textnormal{L}}$(U)  and the SFR of the hosts were positively correlated. We present the same sample in Figure~\ref{tlu-sfr}, including the BCG data points. We estimate $L_{\textnormal{clusters}}$ using only clusters for which a SED fit has been performed. The values of T$_{\textnormal{L}}$(U), and for two other filters, $FUV$ and $B$, are listed in Table~\ref{table-obs2}. The T$_{\textnormal{L}}$(U) and the $\Sigma_\mathrm{SFR}$ are higher in BCGs. In general, the trend suggests that the fraction of star formation happening in clusters is important and increases as function of the SFR. 

To understand why we observe a much higher fraction of T$_{\textnormal{L}}$(U) in BCGs than in spiral galaxies, we need to look at the different star formation histories in these two classes of hosts and not only at the SFRs.  In general, we observe that the young star clusters in BCGs are preferentially clustered in clumps, in agreement with observations of high correlations in position as function of the age among young systems. Studies of spatial correlation among field stars, associations, and clusters show a higher clustering for younger samples, and a clear smoothing of the older structures \citep{2008MNRAS.391L..93G, 2009MNRAS.392..868B, 2010arXiv1010.1837B}. The star formation in spiral systems has proceeded more or less constantly for a long lapse of time. Therefore cluster disruption has worked in favour of populating the stellar fields which form the bulk of the optical luminosity in these systems. In BCGs, the starburst has been acting for rather short timescales ($\sim 40$ Myr), suggesting that cluster disruption (mostly infant mortality) has not been effective. 

Looking at the values listed in Table~\ref{table-obs2}, we see that the fraction of light produced by the star clusters increases at shorter wavelengths and contributes significantly to the UV and U luminosities of the BCGs. Therefore, it suggests that clustered regions in galaxies at redshift $\sim$2-3, with  metallicities similar to the BCGs, contribute to a considerable fraction of the UV-rest frame light. In studies of Lyman break galaxy analogs at redshift between $\sim$0.1 and 0.2, \citet{2008ApJ...677...37O} observed super starburst compact regions which dominate the UV light of these targets. Similarly compact clumps are also observed at redshift $>$1 galaxies \citep{2009ApJ...692...12E, 2011arXiv1104.0248F}. In our analysis, we find some evidence that those bright areas are probably unresolved star cluster knots recently formed and not dispersed yet by the interaction with the galactic environment, e.g. similarly to what is observed in BCGs and in spiral arms \citep{2005A&A...443...79B, 2006ApJ...644..879E}.

\subsection{Do BCGs follow the M$_V^{\textnormal{brightest}}$-SFR relation?}

The M$_V^{\textnormal{brightest}}$-SFR relation was first noticed by L02, who suggested that the visual magnitude of the most luminous cluster and the total galaxy SFR where correlated. By means of numerical modelling, \citet{2004MNRAS.350.1503W} investigated the cluster physical conditions required to reproduce the relation, assuming that the brightest cluster would be also the most massive.  They reproduced the observed relation under the assumption that a cluster population is formed within a time scale of $\sim 10$ Myr, following a CMF {with index steeper ($\sim -2.3$) than is normally observed in cluster populations \citep[e.g. $-2.0$,][]{1999ApJ...527L..81Z, 2003A&A...397..473B}. B08, releasing the condition of the brightest cluster  also being the most massive, observed that the relation showed less scatter if the brightest youngest ($< 10$ Myr) cluster was used instead. Monte Carlo simulations of cluster populations with a CMF of power-law $-2.0$ or a Schechter CMF with power-law $-2.0$ and a characteristic mass of a few times $10^5 \msun$ where used to test different scenarios of cluster formation efficiency. Using the plot shown here in Figure~\ref{mv-sfr}, B08 ruled out a scenario where 100 \% of the stars form in clusters with a CMF power-law of index $-2.0$. The L02 relation appeared to be in better agreement if clusters form with a Schechter CMF and only $\sim 8-10$ \% of the stars reside in bound clusters (cluster formation efficiency, $\Gamma \sim 0.1$). We include the BCG sample in the M$_V^{\textnormal{brightest}}$-SFR, using the corresponding youngest brightest clusters. The 5 targets are located above the relation. The most nearby BCG in our sample, ESO 338, is the one with the largest scatter. In agreement with the M$_V^{\textnormal{brightest}}$-SFR relation, a higher SFR in BCGs enables the formation of more massive (luminous) clusters than quiescent spiral galaxies. Moreover, the position of the targets suggests that the $\Gamma$ in BCGs is higher than the $\sim 8-10$ \% found by B08. 

\begin{figure}
\resizebox{\hsize}{!}{\rotatebox{0}{\includegraphics{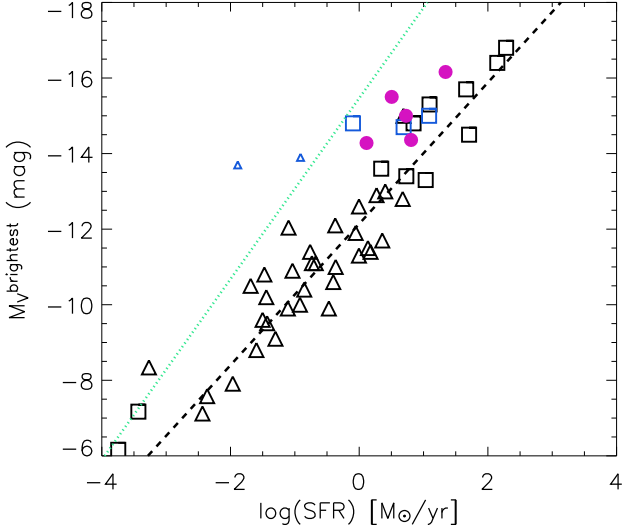}}}
\caption{M$_V^{\textnormal{brightest}}$-SFR relation by \citet{2008MNRAS.390..759B}. The sample from L02 is represented by triangles. Black triangles are the data used to produced the relation indicated by a black dashed line in \citet{2004MNRAS.350.1503W}. Blue triangles are outliers (see L02 and B08). The squares are new data included by B08. In blue we differentiate the  dwarf starburst systems included in the B08 sample. The BCGs are indicated by purple dots. The green dotted line is from B08 and shows where the galaxies should be if all their star formation happens in clusters with a power-law CMF of index $-2.0$.}
\label{mv-sfr}
\end{figure}

A higher CFE in BCGs has already been observed using the G10 relation (\citealp{A2010}; \citealp{A2010c}; \citealp{A2011a}). In Figure~\ref{sfr-cfr}, we show the relation including also SBS 0335 and ESO 338. The position of Haro 11 and ESO 185 has slightly changed compared to the previous publications (\citealp{A2010} and \citealp{A2010c}), because of the different IMF (Salpeter instead of Kroupa) and the different method used to estimate the area of normalization of the SFR (Section \ref{area}). The data by Silva-Villa \& Larsen are clearly scattered around the relation. BCGs follow the trend predicted by the relation, even if  we notice that 3 of the 5 targets sit above the expected values. The mean value of the cluster formation efficiency in BCGs is $\Gamma_{\textnormal{BCGs}}=40 \pm 10$ \%. 

\begin{figure}
\resizebox{\hsize}{!}{\rotatebox{0}{\includegraphics{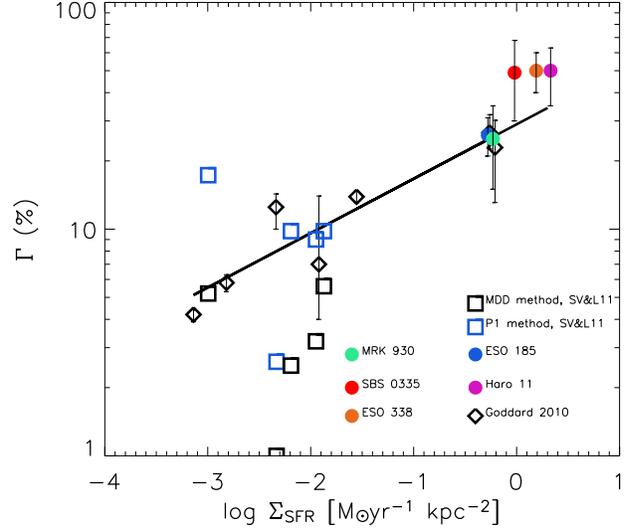}}}
\caption{$\Gamma$-$\log(\Sigma_\mathrm{SFR})$ relation by \citet{2010MNRAS.405..857G}. The sample and the derived relation from Goddard et al. are shown in black diamonds and solid line, respectively. The values estimated by \citet{2011arXiv1101.4021S} are also included as black and blue squares. "P1 method" is similar to the one used by G10 and in this work with the difference that they used clusters with ages between 10 and 100 Myr to constrain the CFR and, thus, $\Gamma$. In the "MDD method" the CFR is estimated from the fit to the observed cluster luminosity function in the galaxy, assuming a disruption model (mass dependent disruption -- MDD) to account for missing clusters. The BCGs are represented by dots (see inset).}
\label{sfr-cfr}
\end{figure}

\section{Cluster formation in BCGs: a close look at high cluster formation efficiencies}

We have used the cluster-host relations to investigate in more detail difference/analogies between BCGs and other star forming galaxies. Due the uncertainties affecting the data we consider the derived quantities as upper limits to the real values. 

One difference of the sample of dwarf galaxies studied by \citet{2002AJ....123.1454B},  luminous BCGs appear to be more efficient in forming star clusters. There are a few possibilities why this may be true. The Billet at al galaxies do not show any clear signature of merging events. Most likely cluster formation in these dwarf systems has been triggered by internal instabilities, low shear, and inflows of gas. Moreover, the gravitational potential in these systems is not deep enough to retain the gas which is clearly observed in outflows. Luminous BCGs have recently experienced a merger event or accreted a considerable amount of gas \citep[see][]{2001A&A...374..800O, A2010, A2010c, A2011a}. This may have favoured the formation of hundreds of massive clusters. Moreover, these systems are likely one or two orders of magnitude more massive than the ones in the Billett et al. sample, allowing them to retain the gas more efficiently. 

The M$_V^{\textnormal{brightest}}$-SFR relation, presented in the previous section, suggests that BCGs are, in the local universe, among the systems which form very massive (bright) clusters. Although they follow the trend it is also clear that they are slightly offset from the prediction made by B08 of a constant cluster formation efficiency. The systematic scatter observed in the position of the BCGs in the M$_V^{\textnormal{brightest}}$-SFR relation suggests that a higher CFE is operating in these systems, which are not dynamically relaxed, e.g., in a merging/interaction phase. The trends observed in the other two relations, $\Gamma$-$\log(\Sigma_\mathrm{SFR})$ and T$_L$(U)-$\Sigma_\mathrm{SFR}$, even if uncertain, support this scenario. A close look at the Goddard et al. relation reveals an interesting point. In the diagram in Figure~\ref{sfr-cfr},  two different groups can be delineated by the distributions of the data points. Using the values listed in Table 4 of G10, we estimate that the mean CFE of the group with lower values (the Silva-Villa \& Larsen sample has been excluded) of $\Sigma_\mathrm{SFR}$ is  $\Gamma=8.7 \pm 4.3$ \%, close to the value found by B08. The more efficient sample includes 2 targets from G10  (the starburst nucleus of the spiral M83 and the LIRG NCG 3256) and the BCGs. Their mean CFE is  $\Gamma = 35.6 \pm 10$ \%, roughly a factor of 4 higher. The two subsamples suggest that cluster formation efficiency is not constant at all scales of SFR. We see that in star-forming systems where star formation has proceeded more or less constantly without any significant burst, the CFE has a mean value of $\sim 8$ \%. Starburst systems, on the other hand, are very active in producing star clusters. Possibly, the difference in CFEs reflects a difference in the conditions of the interstellar medium in the hosts \citep{2008ApJ...672.1006E}. 

The universal cluster formation efficiency discussed by B08 could be valid in the local universe assuming that the interstellar medium density is lower than in high redshift galaxies. However, it is difficult to extend its validity to galaxies with an extreme environment (merging systems). Numerical simulations have shown that GCs may have formed in strongly shocked media and high pressure fields, which have enhanced the gas compression and favoured the formation of more tightly bound structures (\citealp{1997ApJ...480..235E}, \citealp{2008MNRAS.389L...8B}). Such conditions are usually reached in galaxy mergers, where the very massive young star clusters are observed (Antennae system, \citealp{2005A&A...443...41M}; Arp 220, \citealp{2006ApJ...641..763W}; the Bird galaxy, \citealp{2008MNRAS.384..886V}). An increase of the CFE as function of redshift has also been found in simulations by \citet{2010ApJ...718.1266M}. They observed that the cluster formation efficiency (e.g. the mass in cluster versus the total mass of the host) at redshift $\sim 3$ was much higher (about 20 \%) than then in the local universe. For this reason it is possible that the formation and survival of  young and massive clusters formed in extreme environments such as BCGs or LIRGs could help to trace the formation of the old GCs, if they have formed under similar conditions.

 However, only better data in terms of number statistics and quality (high resolution is required in order to distinguish clusters from associations) are needed to reach more stringent conclusions.

\section{Conclusions}

To understand the role of BCGs in the galaxy formation scenario, we have tried to constrain how the star and cluster formation is operating in these systems.

We have looked into the efficiency of the formation of massive star clusters. It is known that  some properties of the cluster population and the mean star formation in the host system are correlated, suggesting that the formation of a cluster is not a local event but intrinsically connected to the mean properties of the galaxy. In general, we find that BCGs follow fairly well these relations, even if their SFRs and cluster properties are more extreme.

We discuss possible uncertainties affecting our results. The relation which is least affected is the M$_V^{\textnormal{brightest}}$-SFR one, which suggests that the cluster formation efficiency  is higher in BCGs than in quiescent spiral and dwarf starburst galaxies. The same evidence is also suggested by the other two relations, $\Gamma$-$\log(\Sigma_\mathrm{SFR})$ and T$_L$(U)-$\Sigma_\mathrm{SFR}$, which appear to be consistent with the general picture despite the uncertainties of the data. 

In particular, we observe that the inclusion of the BCG sample separates the $\Gamma$-$\log(\Sigma_\mathrm{SFR})$ plane into two regions. Local spiral galaxies and dwarf starbursts as the Magellanic Clouds occupy the area around a mean CFE of  $8 \pm 5$ \% in agreement with the prediction made by B08. The BCGs, together with a LIRG and a nuclear starburst region (included in the G10 sample), are in a region with a mean CFE of $35.6 \pm 10$ \%. This indicates that the merger event has enhanced the cluster formation in these systems. 

We observe that the fraction of light produced by  star clusters in BCGs increases at shorter wavelengths and contribute significantly to the UV and U luminosities. This suggests that clustered regions contribute a substantial fraction to the rest-frame UV light of z$\sim$1-3 galaxies with  metallicities similar to the BCGs. In studies of Lyman break galaxy analogs at redshift between $\sim$0.1 and 0.2, \citet{2008ApJ...677...37O} observed super starburst compact regions which dominate the UV light of these targets.  In our analysis, we find some evidence that those bright clumps are probably unresolved star cluster knots, similarly to the ones observed in BCGs.

\section*{Acknowledgments}
A.A. thanks Ana Chies-Santos, for reading the manuscript and interesting discussions on globular cluster formation scenarios, and Esteban Silva-Villa for useful inputs on the interpretation of the current host-cluster relations. The referee, Nate Bastian is thanked for the numerous suggestions and valuable comments which have improved this work.  G.\"O. is a Royal Swedish Academy of Sciences research fellow, supported from a grant from the Knut and Alice Wallenberg foundation. A.A, G.\"O. and E.Z. also acknowledge support from the Swedish Research council and the Swedish National Space Board. This research has made use of the NASA/IPAC Extragalactic Database (NED) which is operated by the Jet Propulsion Laboratory, California Institute of Technology, under contract with the National Aeronautics and Space Administration.

\appendix

\bsp

\label{lastpage}

\end{document}